\documentclass{article}
\usepackage{spconf}

\usepackage{blindtext}
\usepackage{cite}
\usepackage{algorithm,algorithmicx}
\usepackage{algpseudocode}
\usepackage[small,bf]{caption}
\usepackage[cmex10]{amsmath}
\usepackage{amssymb,latexsym,epsfig,subfigure,epic,amscd,mathrsfs,euscript,eufrak,bm}
\usepackage{color}
\usepackage{booktabs}
\usepackage{array}
\usepackage{comment}
\usepackage{epstopdf}
\usepackage{amsfonts}
\usepackage{amsopn}
\usepackage{graphicx}
\usepackage{url} 
\usepackage{empheq}
\usepackage[dvipsnames]{xcolor}

\DeclareMathOperator{\diag}{diag}
\newcommand{\Def}[0]{\mathrel{\mathop:}=}
\newtheorem{myprop}{\bf{Proposition}}

\DeclareMathOperator*{\minimize}{\text{minimize}}

\DeclareMathOperator*{\st}{\text{subject to}}

\title{First-order bifurcation detection  for dynamic complex networks}

\name{Sijia Liu$^{\dagger,*}$, \quad  Pin-Yu Chen$^\S$,   \quad  Indika Rajapakse$^{\ddagger}$,   \quad Alfred Hero$^{\dagger}$
\thanks{This work is supported, in part, by DOE grant DE-NA0002534, the DARPA Biochronicity Program and the DARPA Deep-Purple and FunCC Program.}
}
\address{
$^{\dagger}$Department of Electrical Engineering and Computer Science\\
$^{\ddagger}$Department of Computational Medicine \& Bioinformatics\\
University of Michigan, Ann Arbor, MI 48109, USA\\
$^{\S}$IBM Research AI, Yorktown Heights, NY 10598, USA\\
$^{*}$MIT-IBM Watson AI Lab, IBM Research, Cambridge, MA 02142, USA\\
$^{*,\S}$\{sijia.liu, pin-yu.chen\}@ibm.com~~~
$^{\dagger,\ddagger}$\{indikar, hero\}@umich.edu
}

\begin{document}
%
\maketitle
\begin{abstract}
In this paper, we explore how  network centrality and network entropy can be used to identify a bifurcation network event. 
A bifurcation often occurs when a network   undergoes a qualitative change in its structure  as a response to internal changes or  external signals.
In this paper, we show that 
 network centrality allows us to capture important topological properties of dynamic networks. 
By extracting multiple centrality features from a  network for dimensionality
reduction, we are able to track the  network dynamics underlying an intrinsic low-dimensional manifold. Moreover, we employ von Neumann graph entropy (VNGE) to measure the  information divergence   between networks over time. 
In particular, we   propose an
asymptotically consistent estimator of VNGE so that 
the     cubic complexity of VNGE is reduced to quadratic complexity that   scales more gracefully with network size. 
Finally,  the effectiveness of our approaches is demonstrated through   a real-life application of  cyber intrusion detection. 
\end{abstract}
\section{Introduction}
\label{sec:intro}
Many real-world  complex systems 
  ranging from physical systems, social media, financial markets and ecosystems    to chemical reaction mechanisms  are often represented as networks (or graphs) that possibly change over time \cite{newman2011structure,bertrand2013seeing,sanmou14_mag,chen2010information}.   In a network representation,
a set of elementary units, such as human, gene, sensor,  or other types of `nodes\rq{},   are connected by `edges\rq{} that describe relationships between nodes   such as physical link, spatial vicinity,  or friendship.
Network representations allow us to explore   structural properties of  dynamic systems, and thus to 
 study their   behaviors, e.g., anomaly detection in cyber networks and community detection in social networks \cite{akoglu2015graph,chen2015deep}. 
In  a dynamic system, there often exists  a  critical time instant  at which the system shifts abruptly from one state to another.  This critical threshold is associated with a first-order bifurcation \cite{borchert1981bifurcation} that occurs
 when a small  change made to the  system results in  a sudden  change of the system\rq{}s behavior.
For example, a bifurcation of biological system was detected in the process of cell development when cells
choose between two different fates \cite{liu2017genome,spencer2013proliferation,bargaje2017cell,scheffer2009early}.
In this paper, we aim to explore how 
 network-based approaches can be used in bifurcation detection. 

Centrality measures   provide important means of understanding  the topological structure and dynamic process of complex networks \cite{wang2017identifying}.
Depending on the type of nodal influence to be emphasized,
different centrality measures, such
as degree, eigenvector, clustering coefficient, closeness and betweenness, are commonly used in the literature \cite{newman2011structure}. 
For example, degree centrality measures the total number of connections a node has, while eigenvector centrality implicitly 
measures the importance of a node by the importance of its
 neighbors.  
Network centrality allows us to  
 capture multiple structural features from a single network, and  thus expands the feature set for graph learning under  limited network data samples.
In this paper, we propose a spectral decomposition approach   that integrates multiple network centrality features for graph learning.
It is worth mentioning that our work is different from graph principal
components analysis (PCA) \cite{jiang2013graph,shahid2015robust}, where a graph Laplacian matrix was  used to construct a smooth regularization
function in PCA by assuming  that the data manifold is encoded in the graph.
In contrast  to graph PCA, our approach finds the intrinsic low-dimensional manifold  embedded in the centrality features. 
We    show in this paper that the use of network centrality   helps  to identify differences
in temporal networks.

In addition to network centrality, we employ von Neumann graph entropy (VNGE) to quantify the
  intrinsic network complexity. VNGE  was originally introduced by \cite{braunstein2006laplacian},  determined by the spectrum of the graph Laplacian matrix. It was shown in \cite{anand2011shannon,han2012graph,passerini2008neumann} that VNGE can 
measure the amount of information encoded
in structural features of networks.   For example,  the entropy of random networks, e.g., Erd\H{o}s-R\'{e}nyi random graphs, is larger than
the entropy of scale-free networks under the same average nodal degree
\cite{passerini2008neumann,du2010note}.
Compared to the existing graph entropy measures   using
  notions of either randomness complexity or statistical
complexity \cite{torsello2006learning,feldman1998measures}, the main advantage of VNGE is its computational efficiency, leading to $O(n^3)$ complexity in which $n$ is the network size. Here, we further improve the cubic complexity to $O(n^2)$ by deriving a quadratic approximation to the VNGE. We show that such an approximation is asymptotically consistent, converging to the VNGE under mild conditions. 
Our experiments on a real-life application, cyber intrusion detection, demonstrate that the proposed approaches can efficiently track    the     structural changes of dynamic networks 
 and identify   bifurcation   events. 

\section{Preliminaries: Graph Representation}
\label{sec: reprog}
A graph yields a succinct representation of interactions
among   nodes. 
Mathematically, 
we denote by $\mathcal G = (\mathcal V, \mathcal E, \mathbf W )$ an undirected weighted graph, where
$\mathcal V$ and $\mathcal E$
denote the node and edge sets with cardinality
$|\mathcal V| = n$ and $|\mathcal E| = m$, and $\mathbf W \in \mathbb R^{n \times n}$ is a  weighted matrix with entry $W_{ij}$ (or $[\mathbf W]_{ij}$) satisfying $W_{ij} = 0$ if  $i=j$ or $(i,j) \notin \mathcal E$.
The quantitative study of $\mathcal G$ is often performed under its graph  Laplacian matrix
   $\mathbf L = \mathbf D - \mathbf W$, where
  $\mathbf D = \diag (\mathbf W \mathbf 1)$ is the degree matrix. Here
 $\diag( \mathbf a )$ denotes  the diagonal matrix with    diagonal vector $\mathbf a$, and $\mathbf 1$ is the vector of all ones.
 It is known from
   spectral graph theory \cite{chung1997spectral} that
the second smallest eigenvalue of $\mathbf L$, called {Fiedler number} (FN),  measures the network connectivity. And the number of zero eigenvalues of $\mathbf L$ gives the number of connected components of $\mathcal G$.
Using the above notation, a dynamic network  in a period of length $T$ can be represented as a sequence of graphs $\{ \mathcal G_t \}_{t=1}^T$, where $ \mathcal G_t  = (\mathcal V_t, \mathcal E_t, \mathbf W_t )$. Throughout the paper, we assume that the dynamic network  contains the same set of nodes with $|\mathcal V_t| = |\mathcal V| = n$ for any $t$.


\section{Network Diagnostics via Centrality Analysis}
\label{sec: centrality}

In this section,
 we introduce a graph diagnostic 
method that combines  multiple  centrality features to evaluate nodal importance to the network structure.
By decomposing
a single graph into multiple centrality features, we are able to achieve dimensionality
reduction and feature decorrelation of the graph.
We   introduce several centrality measures of $\mathcal G$ that will be used in the sequel to define our feature set. 

{\textit{Degree (Deg)}}  of  node $i$ is defined as
 $
 \mathrm{Deg}(i) = L_{ii}=\sum_{j=1}^n W_{ij}
 $, where $\mathbf L$ is the graph Laplacian matrix.  
 
{\textit{Eigenvector centrality (Eig)}}
 is defined as
  the  eigenvector of the adjacency matrix $\mathbf W$ associated with its largest positive 
eigenvalue $\lambda_{\max}$. 
The eigenvector centrality of   node $i$ is given by
$
\mathrm{Eig}(i) = [\mathbf v]_i
$ satisfying
$\lambda_{\max} \mathbf v = \mathbf W \mathbf v$.
Eigenvector centrality measures the importance of a node by the importance of its neighbors \cite{newman2011structure}.

{\textit{Local Fiedler vector centrality (LFVC)}}  evaluates the  impact of node removal on  the network connectivity and partition  \cite{chen2015deep}. LFVC of node $i$ is given by
$
\mathrm{LFVC}(i) = \sum_{j \in \{ j | (i,j) \in \mathcal E \}} ( [\mathbf f]_i - [\mathbf f]_j )^2$,
where $\mathbf f$ is the eigenvector (known as
  Fiedler vector) of $\mathbf L$ associated with the smallest non-zero eigenvalue.

{\textit{Closeness (Clos)}}  is a global measure of geodesic
distance of a node to all other nodes \cite{sabidussi1966centrality}. 
Let $\rho(i,j)$ denote the shortest path distance between node $i$ and node $j$ in a connected network. The closeness of node $i$ is defined as
$
\mathrm{Clos}(i) = \frac{1}{\sum_{j \in \mathcal V, j\neq i} \rho(i,j) }$.

{\textit{Betweenness (Betw)}}
measures the fraction
of shortest paths passing through a node relative to
the total number of shortest paths in the network \cite{freeman1977set}.
The betweenness of node $i$  is defined as
$
 \mathrm{Betw}(i) = 
 \sum_{{ l \neq i}}
   \sum_{{ j \neq i, j > l}} \frac{\phi_{lj}(i)}{\phi_{lj}}
$,
 where $\phi_{lj}$ is the total number of shortest paths from node $l$ to $j$, and $\phi_{lj}(i)$ is the number of such shortest paths passing through node $i$.

{\textit{Local clustering coefficient (LCC)}}  quantifies how close a node's neighbors are to become a  complete graph \cite{watts1998collective}. 
LCC of node $i$ is given by 
$
\mathrm{LCC}(i) = \frac{ | \{ (j,l) | j  \in \mathcal N_{i} ,  l  \in \mathcal N_{i}, (j,l) \in \mathcal E \} |}{| \mathcal N_i  | ( | \mathcal N_i  |  - 1)/2}
$,
where $\mathcal N_i $ is the   neighborhood set of node $i$. 

 {\textit{Other topological features:}}
The set of  {hop walk statistics} is another useful network feature that takes into account indirect interactions among nodes. A node's $h$-hop walk weight   is given by the sum of   edge weights associated with   paths     departing from this node and traversing through $h$ edges \cite{chen2016multi}. Moreover, given a set of reference 
    nodes of interest, one can further expand the   feature set by computing graph distances from     reference nodes to other nodes   \cite{liu2017genome}.

Let $\mathbf X_t \in \mathbb R^{n \times p}$  denote the     centrality-based feature matrix 
for a  network at time $t$,   where $n$ is the graph size, and $p$ is the number of centrality features. \textcolor{black}{
In contrast to  graph PCA methods, which are often limited
to    undirected and connected graphs, our approach can be applied to directed and disconnected graphs. This is due to the fact that centrality features are also defined for directed and disconnected graphs.}  After acquiring the feature matrix
$ \mathbf X_t $, 
both linear and non-linear dimensionality reduction techniques   \cite{van2009dimensionality} can be applied. As a result, we obtain a  low-dimensional data representation $\mathbf Y_t \in \mathbb R^{n \times l}$ with  $l \leq p$. 

To better track the   state of a dynamic network, we       fit  the data $\mathbf Y_t$  to  a  minimum  volume  ellipsoid  (MVE), 
 representing a certain confidence region for the state 
\cite{van2009minimum}. 
   The MVE estimate  at time $t$ can then be acquired by solving the   convex program
\begin{align}\label{eq: mve}
\begin{array}{cl}
\displaystyle \minimize_{\mathbf Q \in \mathbb R^{l \times l}, \mathbf b \in \mathbb R^l} & \det(\mathbf Q^{-1}) \\
\st & \|  \mathbf Q \mathbf y_{i,t}  - \mathbf b \|_2 \leq 1, ~ i \in \mathcal N_\alpha \\
& \mathbf Q \succ 0,
\end{array}
\end{align}
\textcolor{black}{where $\mathbf Q$ and $\mathbf b$ are optimization variables, 
$\mathbf y_{i,t}^T$ denotes the $i$th row   of $\mathbf Y_t$, and
$\mathcal N_\alpha$ denotes the set of data within a $\alpha$ confidence region,   determined by
Mahalanobis distances of data  below  
 $\alpha = 97.5\%$ quantile of the chi-square distribution with  $l$ degrees of freedom \cite{van2009minimum}.
 The rationale behind 
problem \eqref{eq: mve} is that 
$\mathbf P \Def \mathbf Q^2$ and $\mathbf c \Def \mathbf Q^{-1} \mathbf b$ defines the ellipsoid 
$
\{ \mathbf x \in \mathbb R^l | (\mathbf x - \mathbf c)^T \mathbf P (\mathbf x - \mathbf c) \leq 1 \}
$, where  the determinant  of $\mathbf P$ (or $\mathbf Q$) is inversely proportional to the volume of this ellipsoid \cite{sun2004computation}.}  Since problem \eqref{eq: mve} involves a linear matrix inequality,
it can be  solved via semidefinite programming (SDP), e.g., the sdp solver in CVX \cite{grant2008cvx}.  
The complexity of  SDP
is approximately of order   $O(a^2 b^{2} + a b^{3})$ \cite{nem12}, where $a$ and $b$
denote the number of optimization variables and the size of
the semidefinite matrix, respectively. In \eqref{eq: mve}, we have $a = l + (l+1)l/2$ and $b = l$,  yielding the complexity $O(l^6)$.  Thanks to dimension reduction,  problem \eqref{eq: mve} can be efficiently solved under a reduced feature space with small $l$.

 \section{von Neumann graph entropy
 }

Von Neumann  entropy (or quantum entropy) was originally used to measure 
    the in-compressible information content
 of a quantum source, 
 and can 
 characterize   
  the departure of a dynamical system from a pure state with zero entropy \cite{anand2011shannon}.  
Recently,  the von Neumann entropy of a graph, known as von Neumann graph entropy (VNGE),  was used to efficiently measure the graph complexity \cite{han2012graph}. 
By constructing a scaled
   graph Laplacian matrix  $\mathbf L_c \Def c \mathbf L$ with $c = 1/\mathrm{trace}(\mathbf L)$,
VNGE can be defined as \cite{anand2011shannon,passerini2008neumann}
 \begin{align}\label{eq: VNGE}
V = - \sum_{i=1}^n \lambda_i \log \lambda_i, 
 \end{align}
 where $\mathrm{trace}(\cdot)$ denotes the trace operator of a matrix,  
$\{ \lambda_i \}_{i=1}^n$ are eigenvalues of 
 $\mathbf L_c$, and the convention
$0 \log 0 = 0$ is used since $\lim_{x \to 0^+} x \log x = 0$. 
It is clear from \eqref{eq: VNGE} that
 VNGE can be interpreted as the Shannon entropy of the probability distribution  represented by  $\{ \lambda_i \}_{i=1}^n$ under the conditions that $\lambda_i \geq 0$ for any $i$ and $\sum_{i=1}^n \lambda_i = 1$.  
Therefore, 
regular graphs with an uniform distribution of $\{ \lambda_i \}_{i=1}^n$ provides an upper bound on VNGE. It was proved by \cite{du2010note} that the
VNGE 
of  Erd\H{o}s-R\'{e}nyi random graphs saturates this upper bound.


In \eqref{eq: VNGE},
VNGE
requires the full eigenspectrum of the graph Laplacian matrix, and thus has 
the cubic computational complexity $O(n^3)$.
In Proposition\,\ref{prop: VNGE}, we propose an approximate VNGE that scale more gracefully with the network size $n$. 
\begin{myprop}\label{prop: VNGE}
The quadratic approximation $\mathbf Q$ of VNGE $\mathbf V$ in \eqref{eq: VNGE} is given by
 \begin{align}\label{eq: VNGE_approx}
 Q =  1 - c^2 \left ( \mathbf d^T \mathbf d + \mathbf 1^T ( \mathbf W \circ \mathbf W )  \mathbf 1 \right ),
 \end{align}
where $\mathbf d$ is the diagonal vector of $\mathbf L$, $\mathbf W$ is the weighted adjacency matrix, 
and $\circ$ denotes the entrywise product. Moreover, 
\[
Q \to \frac{V}{\log{n}}, \quad n \to \infty,
\]
when $n_+ \sim n$ and $\lambda_{\max} \sim \lambda_{\min}$, where
$n_+$ denotes the number of positive
eigenvalues of $\mathbf L_c $,  
$\lambda_{\max}$ and $\lambda_{\min}$ denote the largest and smallest nonzero eigenvalues of $\mathbf L_c$, and for 
two  functions $f(n)$ and $g(n) \neq 0$, $f(n) \sim g(n)$ means $\lim_{n \to \infty} f(n)/g(n) = 1$.
\end{myprop}
\textbf{Proof:} See Proof in Sec.\,\ref{sec: app}. \hfill  $\blacksquare$

In contrast with \eqref{eq: VNGE}, the quadratic approximation \eqref{eq: VNGE_approx} yields an improved  computational complexity of order $O(n^2)$.  Moreover, Proposition\,\ref{prop: VNGE} implies that the asymptotic consistency of $Q$ with respect to $V$ is guaranteed up to a constant factor $\log{n}$. The condition $n_+ \sim n$ implies that the number of disconnected components (given by $n - n_+$) is  ultimately negligible compared to $n$. The condition $\lambda_{\max} \sim \lambda_{\min}$ implies that a graph Laplacian matrix has balanced eigenspectrum. 
This condition    holds
in regular and homogeneous random graph  \cite{passerini2008neumann}.

 

\section{Experimental Results}
\label{sec: bifur}

   \begin{figure}  
  \centering
 \hspace*{-0.1in}\begin{tabular}{c}
\includegraphics[width=0.38\textwidth,height=!]{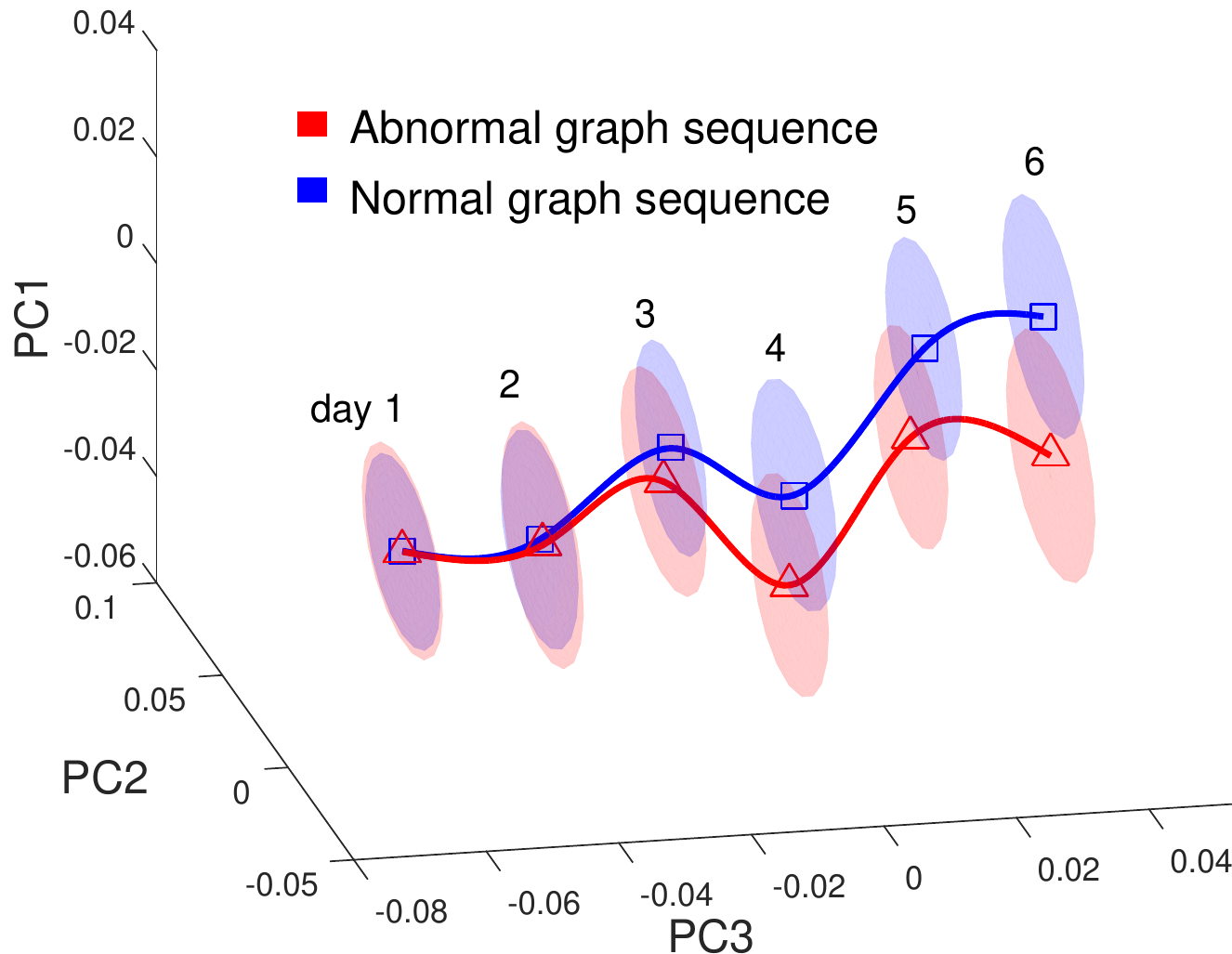} 
\\
(a)  \\
\hspace*{-0.0in}\includegraphics[width=0.38\textwidth,height=!]{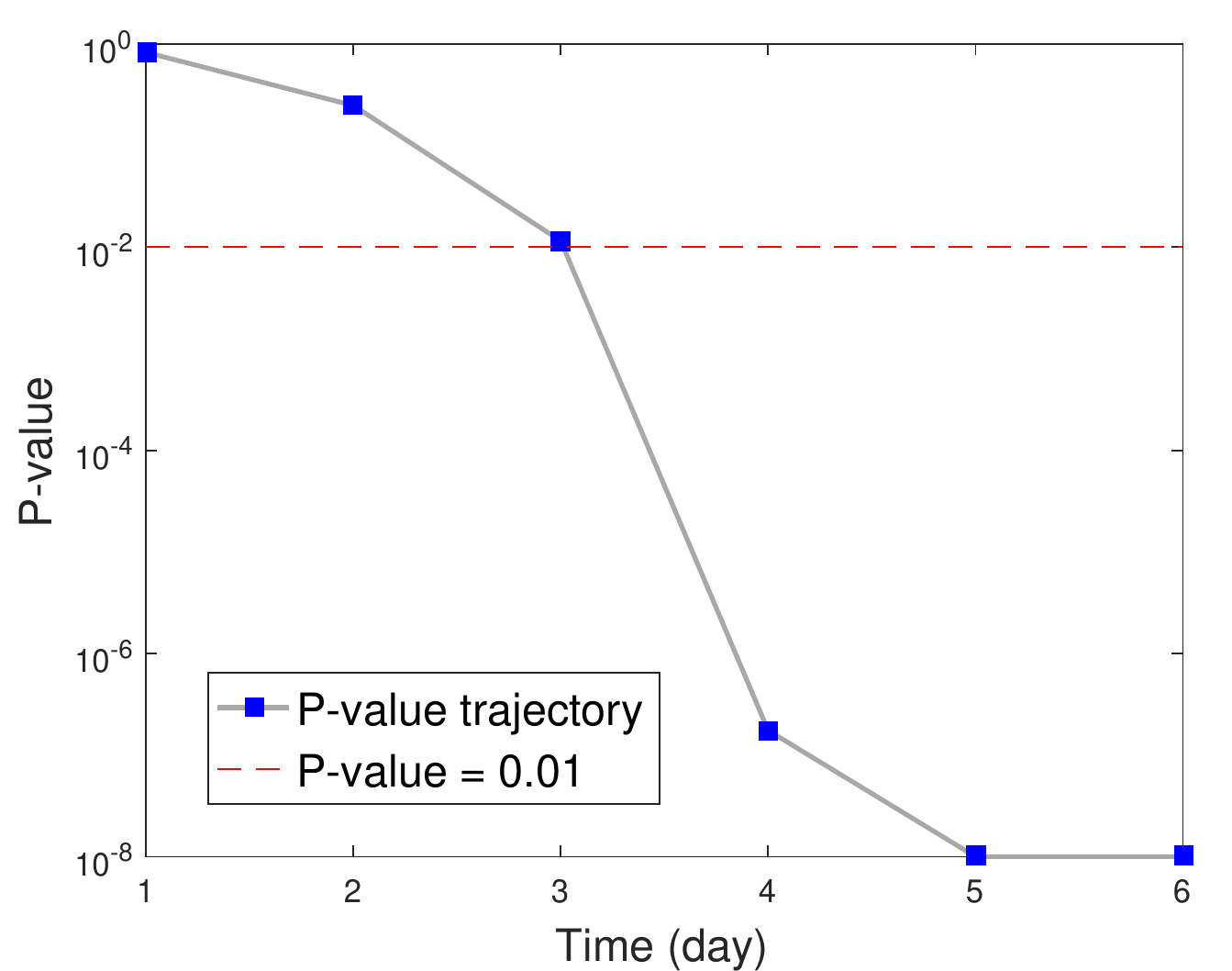} 
\\
(b)
\end{tabular}
 \caption{Centrality-based bifurcation detection. (a) MVEs of  centrality features within $97.5\%$ confidence region. (b) $P$ value vs time. }
\label{fig: bifurcation}
\end{figure}

In this section, we demonstrate the effectiveness of network centrality and VNGE  in first-order bifurcation detection.   We conduct our experiments using the UNB intrusion detection evaluation dataset  \cite{shiravi2012toward,chen2016multi}.  Here  two graph sequences  (with known adjacency matrices)  are given  
in a time period of $6$ days, and each of them corresponds to a 
cyber network in which each
node is a  machine  and an edge implies the
presence of communication between machines. 
 The first graph sequence  describes the normal activity of cyber networks from day $1$ to day $6$. 
The second graph sequence includes abnormal
networks under  denial of service (DoS) and infiltrating attacks
from day $4$ to day $6$. 

In Fig.\,\ref{fig: bifurcation}, we  present principal component analysis based on 
 network centrality features  extracted  from normal and abnormal networks.  
In Fig.\,\ref{fig: bifurcation}-(a), we 
present MVEs that fits the 3D   representations of network centrality features obtained from PCA.  
Here the trajectory of centroids is  smoothed using the cubic spline.  As we can see, there exists a {divergence} between the normal graph sequence and the abnormal graph sequence. The  abrupt change from day $3$ to day $4$ reflects the   anomalous behavior of the network after day $3$  when the attack began. The observed branching trajectory   can be interpreted using the concept of {bifurcation} \cite{borchert1981bifurcation}:
there exists   
a bifurcation   of order $1$
in the sense that 
two simultaneous trends starting from the same type of    networks (non-attacked)  become separated from each other, toward two different types of networks (non-attacked versus attacked). 
 In Fig.\,\ref{fig: bifurcation}-(b), we 
 evaluate the   significance 
of the difference between the normal graph sequence and the abnormal graph sequence. 
\textcolor{black}{Here the  $P$ value
is defined from the  Hotelling's T-squared test  \cite{manly2016multivariate} associated with the null hypothesis that the centroids of  ellipsoids
from the normal and abnormal graph sequences are identical at a given time point.}
 Clearly,  day $4$ is the critical time   for  cyber intrusion
with $P\text{ value} < 0.01$.

      \begin{figure}[htb]   
  \centering
\includegraphics[width=.38\textwidth]{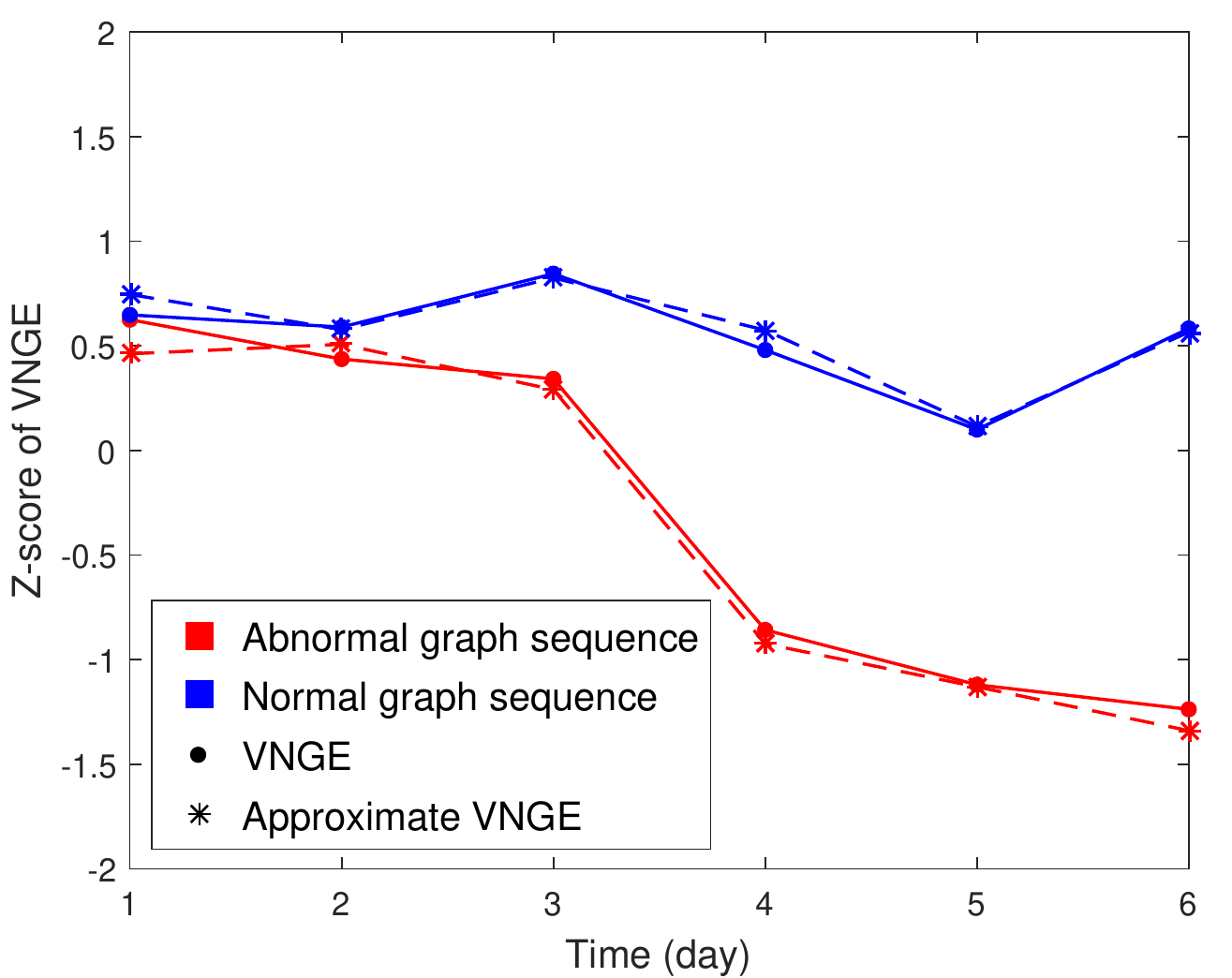}
\caption{{VNGE of cyber networks over time.
}}
\label{fig: VNGE}
\end{figure}

In Fig.\,\ref{fig: VNGE}, we present   VNGE of the studied two graph sequences, where VNGE is shown by its Z-score, which is normalized over time to have zero mean and unit variance. 
 As we can see, the entropic pattern of the normal graph sequence is quite different from that of the abnormal sequence. Similar to Fig.,\ref{fig: bifurcation}, there exists a first-order bifurcation after day $3$.
 An interesting observation is that the VNGE of abnormal networks after bifurcation  is   lower than normal networks prior to bifurcation. That is because
 the network becomes more heterogeneous under attacks, e.g., DoS attack
 is   accomplished by flooding some targeted machines  with superfluous requests in an attempt to overload these hosts. And the VNGE decreases when the degree heterogeneity of a network increases  \cite{passerini2008neumann,anand2011shannon}.  
 We further note that the approximate VNGE  is close to the exact VNGE over time, implying    the effectiveness of our proposed low-complexity approximation given by \eqref{eq: VNGE_approx}. 

\section{Conclusion}
This paper showed how one can use   network centrality and VNGE to detect bifurcation of dynamic complex  networks. 
Network centrality enables us to capture important topological properties of dynamic networks, and VNGE provides an efficient approach to measure the information divergence between dynamic networks.
When applied to cyber
intrusion detection,  our approaches   effectively detected the network bifurcation event. 
 In the future, we would like to delve into the relation between bifurcation and network entropy.
We will also apply our approaches to 
other real-life applications.

\newcommand{\bLN}{\mathbf{L}_{\mathcal{N}}}
\newcommand{\bLNN}{\mathbf{L}_{N}}
\newcommand{\ra}{\rightarrow}

\section{Proof of Proposition\,1}
\label{sec: app}
Based on Taylor series expansion of  $   \log x = \sum_{k=1}^\infty \frac{(-1)^{k-1}}{k}(x-1)^k$ at $x = 1$, we have quadratic approximation 
 \begin{align*} 
V \approx Q = & - \sum_{i=1}^n \lambda_i (\lambda_i - 1) {=} 1 - \sum_{i=1}^n \lambda_i^2 = 1 - \sum_{i=1}^n \sum_{j=1}^n [\mathbf L_c]_{ij}^2 \nonumber \\
\stackrel{(a)}{=} & 1 - c^2 \left (
\sum_{i=1}^n L_{ii}^2 + \sum_{i=1}^n \sum_{j \neq i} L_{ij}^2 \right ) 
=  \text{\eqref{eq: VNGE_approx}},
 \end{align*}
where the equality $(a)$ holds due to the definition of $\mathbf L_c$ in \eqref{eq: VNGE}, and $[\mathbf A]_{ij}$ (or $A_{ij}$) represents the $(i,j)$-th entry of a matrix $\mathbf A$. 
  Assuming  $\mathbf L_c$ has at least two nonzero eigenvalues, which implies $0< \lambda_i \leq \lambda_{\max} < 1$ for any nonzero eigenvalue $\lambda_i$. We rewrite $V$ as 
\begin{align}
\label{eqn_thm_entropy_bound_1_c}
V 
=-\sum_{i: \lambda_i > 0} \lambda_i \ln \lambda_i 
=-\sum_{i: \lambda_i > 0} \lambda_i (1-\lambda_i) \frac{\ln \lambda_i}{1-\lambda_i}.
\end{align}
Since for all $\lambda_i > 0$, $\ln \lambda_{\min} \leq \ln \lambda_i \leq\ln \lambda_{\max}<0$ and $0<1- \lambda_{\max} \leq 1-\lambda_i \leq 1- \lambda_{\min} <1$, we obtain the relation
\begin{align}
\label{eqn_thm_entropy_bound_2}
\frac{-\ln \lambda_{\max}}{1-\lambda_{\min}} \leq \frac{-\ln \lambda_{i}}{1-\lambda_{i}} \leq  \frac{-\ln \lambda_{\min}}{1-\lambda_{\max}}.
\end{align}
Using $Q=\sum_{i=1}^n \lambda_i(1 - \lambda_i)=\sum_{i: \lambda_i>0} \lambda_i(1 - \lambda_i)$  and applying (\ref{eqn_thm_entropy_bound_2})  to (\ref{eqn_thm_entropy_bound_1_c}), we have
\begin{align}
\label{eqn_thm_entropy_bound_3}
-Q \frac{\ln \lambda_{\max}}{1-\lambda_{\min}} \leq V \leq -Q \frac{\ln \lambda_{\min}}{1-\lambda_{\max}}.
\end{align}
Let  $\lambda_{\max}=\frac{a}{n}$ and  $\lambda_{\min}=\frac{b}{n}$ for some constants $a,b$ such that $a \geq b >0$. We obtain
\begin{align*}
\lim_{n \ra \infty}-\frac{1}{\ln n} \cdot \frac{\ln \lambda_{\max}}{1-\lambda_{\min}}= \lim_{n \ra \infty} \frac{1}{\ln n}  \cdot \frac{\ln n - \ln a}{1-\frac{b}{n}} = 1.
\end{align*}
Similarly, 
$
\lim_{n \ra \infty}-\frac{1}{\ln n} \cdot \frac{\ln \lambda_{\min}}{1-\lambda_{\max}} = 1$. 
Taking the limit of $\frac{V}{ \ln n}$ and applying the above results into  (\ref{eqn_thm_entropy_bound_3}), we finally obtain
$
\lim_{n \ra \infty} \frac{V}{\ln n} - Q= 0$.
\hfill $\blacksquare$

%


\vfill\pagebreak

\bibliographystyle{IEEEbib}
{\small{
\bibliography{Bio_ref}
}
}
\end{document}